\documentstyle[12pt,draft]{article}

 \newcommand {\be} {\begin{equation}}
\newcommand {\bea} {\begin{eqnarray} \nonumber }
\newcommand {\ee} {\end{equation}}
\newcommand {\eea} {\end{eqnarray}}

 \newcommand {\lan} {\langle}
 \newcommand {\ran} {\rangle}
\begin{document}   

\title{A Conjecture on random bipartite matching}
\author{ Giorgio Parisi\\
Dipartimento di Fisica, Universit\`a {\em La  Sapienza},\\ 
INFN Sezione di Roma I \\
Piazzale Aldo Moro, Roma 00185, Italy}
\maketitle

\begin{abstract}
In this note we put forward a conjecture on the average optimal length for bipartite matching with a 
finite number of elements where the different lengths are independent one from the others and have an 
exponential distribution.
\end{abstract}
\vfill
%{\bf \hfill ROM}
\vfill

The problem of random bipartite matching (or assignment) is interesting both from the point of view 
of optimisation theory and of statistical mechanics \cite{PS,mp1,mp2,qua}.

Each instance of the problem is characterised by a $N$ cross $N$ matrix $d$; sometimes $d(i,k)$ 
represents the {\sl distance} between $i$ and $k$.  We are interested to compute the length of the 
optimal bipartite matching defined as:
\be
L(d) =\min_\Pi \sum_{i=1,N} d(i,\Pi(i)),
\ee
where the minimum is done over all the $N!$ permutations ($\Pi$) of 
$N$ elements and $\Pi(i)$ denote the results of the action of the permutation $\Pi$ on $i$.

If the matrix $d$ has a given probability distribution, we can 
define the average of $L(d)$ over this probability distribution.  The problem of computing this 
average can be studied using the technique of statistical mechanics.

A well studied case is when the elements of $d$ are statistically independent one from the others.  
Here we can use replica theory \cite{mpv,parisibook2} in the mean field approximation.  When the 
probability distribution of each element of $d$ is flat in the interval $[0,1]$, after a long 
computation it was found
\cite{mp1,mp2} that for large $N$ we have:
\be
\lan L \ran_N = \zeta(2) -{\zeta(2)+2\zeta(3) \over N}
+O\left( {1 \over N^2}\right), \label{REP}
\ee
where $\zeta$ is Riemann zeta function and the brackets denote the average over different instances 
of the system.  This behaviour is well confirmed by numerical simulations
\cite{mp2,qua}.

In this note we consider a slightly different case where the probability distribution of each 
element of $d$ is given by
\be
P(d)= \exp(-d).
\ee

The result in the infinite $N$ limit depends only on the behaviour of 
$P(d)$ around $d=0$, so that
also in this case we expect that
\be
\lim_{N \to \infty} \lan L \ran_N = \zeta(2). \label{LIMIT}
\ee

Here we are interested in an exact formula for $\lan L \ran_N$. 
Generally speaking $\lan L \ran_N$ is
 a rational number.
A simple computation shows that
\be
\lan L \ran_1 =1 \ \ \ \lan L \ran_2= 1+ {1 \over 2^2}.
\ee

Our conjecture is that
\be
\lan L \ran_N = \sum_{k=1,N}{1 \over k^2}.\label{CON}
\ee

It is evident that the exact results for $N=1$ and $N=2$ and the 
limit for $N \to \infty$ (eq.
\ref{LIMIT}) are compatible with our conjecture.

We have tried to falsify numerically our conjecture for $N=3,4,5$ (by randomly sorting $O(10^8)$ 
instances of the matrix $d$).  We have verified that the numerical results are compatible with our 
conjecture with a precision of less that $5\ 10^{-5}$.  It is remarkable that for these value of $N$ 
we have $\lan L^{2}(d)\ran \approx \lan L(d) \ran +1 $.

 If this conjecture (eq.  \ref{CON}) is 
correct, it seems likely that the value $\lan L \ran_N$ can be computed analytically in a compact 
way.  A proof of this result would be valuable also because it would automatically imply the 
correctness of the replica result (eq.  \ref{REP}) in the infinite $N$ limit.  However we have not 
succeeded in this task.

\vskip1cm
It is pleasure for me to thank L.~Cugliandolo,  E.~Marinari and M.~M\'ezard for useful discussions.

 \end{document}